\documentclass[final,1p,times]{elsarticle}
\usepackage{graphicx}
\usepackage{amsmath, amssymb, amsthm, dsfont, bm}         %% math formulae
\usepackage{amsfonts}         %% math fonts
\usepackage{color}
\newcommand {\R} {{\mathds R}}

\newcommand {\C} {{\mathds C}}

\newcommand{\ds}{\displaystyle}
\newcommand{\pl}{\partial }
\newcommand {\TF} {\hbox{\tiny TF}}
\def\Blue#1{{\color{black}{#1}}}
\journal{Journal of Computational Physics}
\begin{document}

\begin{frontmatter}



\title{A finite element method with mesh adaptivity for computing vortex states in fast-rotating Bose-Einstein condensates}


\author[ljll1,ljll2]{Ionut Danaila\corref{io}}
\ead{danaila@ann.jussieu.fr}

\author[ljll1,ljll2]{Fr{\'e}d{\'e}ric Hecht}
\ead{hecht@ann.jussieu.Fr}

\address[ljll1]{UPMC Univ Paris 06, UMR 7598, Laboratoire Jacques-Louis Lions, F-75005, Paris, France}

\address[ljll2]{CNRS, UMR 7598, Laboratoire Jacques-Louis Lions, F-75005, Paris, France}

\cortext[io]{Corresponding author. Tel.: +33-1-44277169;
              fax: +33-1-44277200}

\begin{abstract}
Numerical computations of stationary states of fast-rotating Bose-Einstein condensates  require high spatial resolution due to the presence of a large number of quantized vortices. In this paper we propose a low-order finite element method with mesh adaptivity by metric control, as an alternative approach to the commonly used high order (finite difference or spectral) approximation methods. The mesh adaptivity is used with two different numerical algorithms to compute stationary vortex states: an imaginary time propagation method and a Sobolev gradient descent method. We first address the basic issue of the choice of the variable used to compute new metrics for the mesh adaptivity and show that refinement using simultaneously the real and imaginary part of the solution is successful. Mesh refinement using only the modulus of the solution as adaptivity variable fails for complicated test cases. Then we suggest an optimized algorithm for adapting the mesh during the evolution of the solution towards the equilibrium state. Considerable computational time saving is obtained compared to uniform mesh computations. The new method is applied to compute difficult cases relevant for physical experiments (large nonlinear interaction constant and high rotation rates).
\end{abstract}


\end{frontmatter}



%%%%%%%%%%%%%%%%%%%%%%%%%%%%%%%%%%%%%%%%%%%%%%%%%%%
\section{Introduction}
%%%%%%%%%%%%%%%%%%%%%%%%%%%%%%%%%%%%%%%%%%%%%%%%%%%%

Recent research efforts in the field of condensed matter physics were devoted to the study of quantized vortices nucleated in a Bose-Einstein condensate (BEC). Several groups have produced vortices in different experimental set-ups \cite{matthews,Madison2000a,Madison2000b,Abo-Shaeer2001a,Raman2001,Rosenbusch2002a}, leading to numerous theoretical and numerical studies aimed at a better understanding of such macroscopic superfluid systems with quantized vorticity.

A typical experimental BEC configuration with quantized vortices is the rotating condensate. The condensate is confined by a magnetic potential and set into rotation using a laser beam, which can be assimilated to  a spoon  stirring a cup of tea. Since the solid body rotation is not possible in a superfluid system, the condensate has the choice between staying at rest and rotating by nucleating quantized vortices.
The number and shape of vortices depend on the rotational frequency and the
geometry of the trap.
The fast rotation regime is particularly interesting to explore since a rich
variety of scenarios are theoretically predicted: formation of
giant (multi-quantum) vortices, vortex lattice melting or quantum
Hall effects. This regime is experimentally delicate to
investigate \cite{Rosenbusch2002b,rot-bretin,rot-sabine}, making numerical simulations very appealing in depicting vortex configurations for fast rotations.

\begin{figure}[!h]
\centering
\includegraphics[width=0.8\columnwidth]{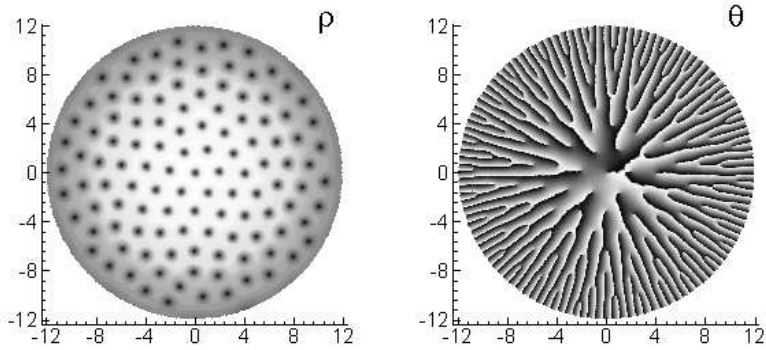}
\caption{Example of fast rotating condensate (harmonic trapping potential, $g=5000$, $\Omega/\omega_\perp=0.95$) computed with the present method. Contours of atomic density $\rho$ (left, low density in black) and phase $\theta$ (right) of the converged (stationary) state. Note the dense Abrikosov  vortex lattice and phase discontinuities joining the border of the condensate.}
\label{fig-abrikosov}
\end{figure}

However, numerical simulations of fast rotating condensates are also very challenging for at least two reasons. The first difficulty comes from the presence in a condensate of a large number of vortices when high rotation frequencies are reached.  An example of such configuration is illustrated in Fig. \ref{fig-abrikosov} for a condensate trapped in a harmonic magnetic potential. We recall that a quantized vortex is a topological defect of the macroscopic wave function describing the condensate:
\begin{equation}\psi=\sqrt{\rho(x,y,z)} \,e^{i \theta(x,y,z)},\end{equation}
 where $\rho$ is
the local atomic density and $\theta$ the phase. In other words, $\rho=0$ in the core of the vortex  (no condensed atoms are present) and around the vortex there
exists a frictionless superfluid flow with a discontinuous  phase
field.
As a consequence of this phase discontinuity, the circulation around a vortex is quantized
\begin{equation}
  \Gamma=\oint {\mathbf v} . {\mathbf dl} = n\frac{h}{m},
\end{equation}
where ${\mathbf v}=\frac{h}{2\pi m} \nabla \theta$ is the local velocity (defined by analogy with classical fluids), $h$ is Planck's constant, $m$ the atomic mass and $n$ an integer (the winding number). A numerical system has to offer sufficient spatial resolution, not only to capture the large gradients of the density $\rho$ in the small-size vortex cores, but also to cope with phase discontinuities that extend up to the edge of the condensate (see Fig. \ref{fig-abrikosov}). This explains the use in the literature of discretization methods with high order spatial accuracy: Fourier spectral \cite{perez-gc,perez-gcM,zeng2009}, sixth-order finite differences \cite{dan-2003-aft,dan-2004-aft,dan-2005}, sine-spectral \cite{bao-norm,bao2006}, Laguerre--Hermite pseudo-spectral \cite{bao2008}, etc.

\begin{figure}[!h]
\centering
\includegraphics[width=0.75\columnwidth,angle=-90]{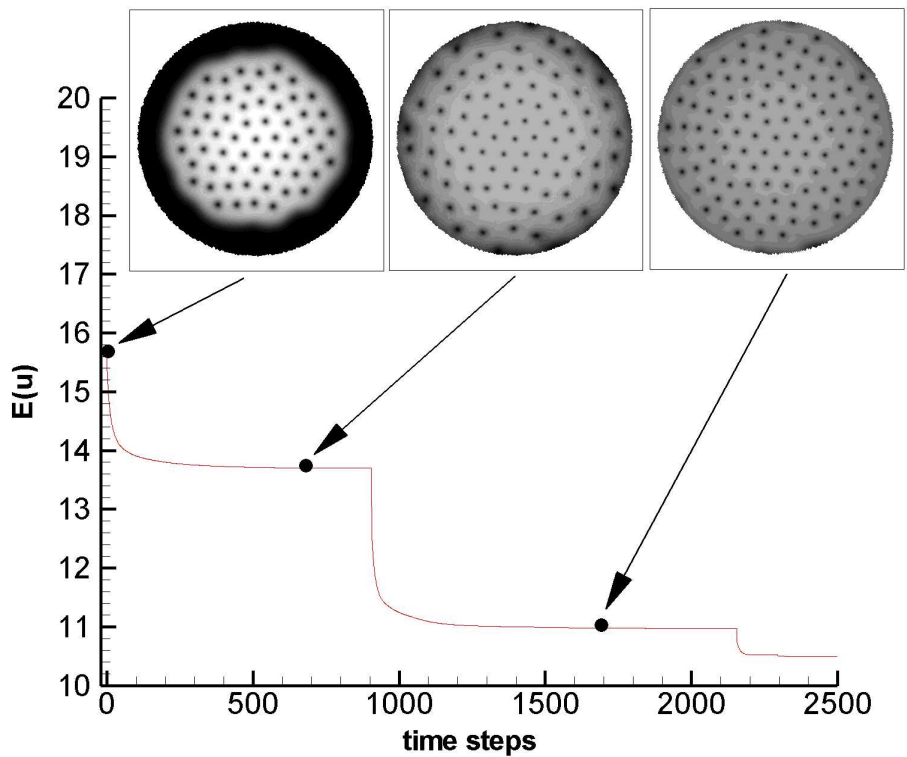}
\caption{Illustration of the imaginary time evolution of the solution before reaching the converged state displayed in Fig. \ref{fig-abrikosov}.
Energy decrease and contours of atomic density $\rho$ for intermediate states.
Note the nucleation of new vortices and their
rearrangement in a more and more regular Abrikosov lattice. The jumps in the energy evolution correspond to mesh refinement.}
\label{fig-time-evol}
\end{figure}

The second numerical difficulty in computing such configurations comes from the numerical algorithm used to converge to stable states with vortices. Most of the numerical algorithms proposed in the literature use the so-called {\em imaginary time propagation} of the wave function. A typical computation (Fig. \ref{fig-time-evol}) starts from an ad-hoc initial configuration  and iteratively search for a minimizer of the energy describing the system (such methods are described in the next section).
During the iterative process, the vortices  move slowly in the condensate towards their final equilibrium locations. Depending on the initial condition, new vortices could also enter the condensate. This is the case in Fig. \ref{fig-time-evol} where a converged computation for a lower value of the nonlinear interaction constant is used as initial condition.
This evolution, called  {\em imaginary time evolution} since it has no physical relevance, has to be accurately captured by the numerical system and brings up the question of the behavior of standard dynamic mesh adaptivity methods in this context. To the best of our knowledge, this question was not addressed in the literature.

In this paper we tackle the two above mentioned difficulties by using a low-order finite element method with mesh adaptivity, as an alternative of commonly used high-order methods. Finite element method have been already used \cite{ADU,bao2003} to compute vortex states in rotating BEC, but with fixed meshes. An attempt to adapt the mesh was made in \cite{baksmaty2009} by using a fixed computational domain with different mesh densities;  finer meshes were initially set in subdomains where vortices are guessed to lie in the final equilibrium configuration.

\Blue{It is important to note that the mesh adaptivity is also of great interest for the simulation of vortex states in type-II superconductors. Such systems are described by the Ginzburg-Landau (GL) macroscopic model that has close resemblance with the GP equation when high values of the GL parameter (kappa) are considered \cite{du-1996-hkappa}.
Several key studies \cite{du-1992-SIAMREV,du-1992-PRB} have set the mathematical and algorithmic basis for the use of finite element method to simulate vortex configurations governed by the GL model (see \cite{du-2005-GL} for a review). However, as mentioned in \cite{du-2005-GL}, using mesh adaptivity in computing vortex lattices with a large number of vortices is still a computational challenge in this field too.
}

The purpose of the present approach is to use a dynamic mesh adaptivity that allows to follow the evolution of vortices during the computation until convergence. To this end, we start by implementing in a low-order (piecewise linear) finite element setting two different algorithms to compute stationary vortex states: a classical method based on the imaginary-time propagation of the wave function and a Sobolev gradient descent method for the direct minimization of the energy functional. These two algorithms are described in the next section. Section \ref{sec-FE} presents the finite element setting and the mesh adaptivity strategy based on  metric control. Several numerical experiments are designed in section \ref{sec-numerics}. We start by
answering the basic question of the choice of the variable used to adapt the mesh. In particular, we show that the approach, that might appear as natural, of refining the mesh following the atomic density $\rho$ is not always successful.  Extensive numerical tests prove that refinement using simultaneously the real and imaginary part of the solution as adaptivity variables is the successful approach. The new adaptive mesh strategy is shown to bring an important computational time saving over computations using refined fixed meshes. Finally, the proposed method is applied to compute difficult cases, with large nonlinear interaction constant and high rotation rates, that are relevant for physical experiments.

%%%%%%%%%%%%%%%%%%%%%%%%%%%%%%%%%%%%%%%%%%%%%%%%%%%
\section{Numerical methods to compute minimizers of the Gross-Pitaevskii energy}\label{sec-GP}
%%%%%%%%%%%%%%%%%%%%%%%%%%%%%%%%%%%%%%%%%%%%%%%%%%%%

%%%%%%%%%%%%%%%%%%%%%%%%%%%%%%%%%%%%%%%%%%%%%%%%%%%%
\subsection{Mathematical problem}
%%%%%%%%%%%%%%%%%%%%%%%%%%%%%%%%%%%%%%%%%%%%%%%%%%%%

In the zero-temperature limit, a  dilute gaseous BEC is mathematically described by a macroscopic complex wave function $\psi(\mathbf{x})$, which spatial configuration  is obtained by minimizing the Gross--Pitaevskii (GP) energy.
We consider a BEC of $N$ atoms trapped in a magnetic potential  $\tilde{V}_{trap}$ with radial symmetry and transverse trapping frequency $\omega_\perp$. The condensate is rotating along the $z$-axis with the angular velocity $\tilde{\Omega}$. It is common practice to scale the variables using as characteristic length the harmonic-oscillator length $d=\sqrt{\frac{\hbar}{m \omega_{\bot}}}$, where $\hbar$ is Planck's constant and $m$ the atomic mass of the gas. Using the scaling,  $\mathbf{r} = \mathbf{x}/d$, $u(\mathbf{r})=\psi(\mathbf{x}) d^{3/2}/\sqrt{N}$, $\Omega = \tilde{\Omega}/\omega_\perp$, we obtain the non-dimensional energy (per particle) in the rotating frame:
\begin{equation}
E(u) = \int_{\cal D} \frac{1}{2} | \nabla u |^2 + V_{trap}|u|^2 + \frac{g}{2}|u|^4 -   \Omega i u^* (A^t \nabla) u,
\label{eq-u-energ}
\end{equation}
where $V_{trap}=\frac{1}{\hbar \omega_{\bot}} \tilde{V}_{trap}$,  and  $A=(y,-x, 0)$. We denote by $u^*$ the complex conjugate of $u$. The interactions between atoms are described by the constant $g=\frac{4 \pi N a_s}{d}$, with $a_s$ the $s$-wave scattering length.
The mass conservation constraint becomes in this scaling:
\begin{equation}
\int_{\cal D} |u|^2 = \| u \|^2 = 1,
\label{eq-u-norm}
\end{equation}
where we denote by $\| . \| = \| . \|_{L^2({\cal D}, \C)}$. Note that we have considered that $u(\mathbf{r}) \rightarrow 0$, as $\mathbf{r} \rightarrow \infty$ and, consequently, the condensate could be confined in a bounded domain ${\cal D}$.

We consider in the following the two-dimensional problem defined on ${\cal D} \subset \R^2$, with homogeneous Dirichlet boundary conditions $u=0$ on $\partial D$. For given constants  $\Omega, g$ and  trapping potential function $V_{trap}$, the minimizer $u_g$ of the functional (\ref{eq-u-energ}) under the constraint (\ref{eq-u-norm}) is called the ground state of the condensate. Local minima of the energy functional with energies larger that $E(u_g)$ are called excited (or metastable)  states of the condensate.

We present in the following two different methods to compute minimizers of the GP energy.

%***************************************************************************************************************************
%%%%%%%%%%%%%%%%%%%%%%%%%%%%%%%%%%%%%%%%%%%%%%%%%%%%%%%%%
\subsection{Imaginary time propagation: Runge-Kutta-Crank-Nicolson scheme}\label{subsec-rk}
%%%%%%%%%%%%%%%%%%%%%%%%%%%%%%%%%%%%%%%%%%%%%%%%%%%%%%%%%

Most of the numerical algorithms proposed in the literature to compute minimizers of the GP energy use
the so-called {\em normalized gradient flow} \cite{bao-norm}. It consists in applying
the steepest descent method for the unconstrained problem,
\begin{equation}
\frac{\partial u}{\partial t} = - \frac{1}{2} \frac{\partial E(u)}{\partial u} = \frac{\nabla^2 u}{2} - V_{trap} u - g  |u|^2u  +   i \Omega A^t \nabla u,
\label{eq-grad-flow}
\end{equation}
\Blue{to advance the solution $u\in \C$ from the discrete time level $t_n$ to $t_{n+1}$; the obtained predictor ${\tilde u}(r,t_{n+1})$ is then normalized
and used to set the solution at $t_{n+1}$ satisfying the unitary norm constraint:
\begin{equation}
u(\mathbf{r},t_{n+1}) \triangleq \frac{{\tilde u}(\mathbf{r},t_{n+1})}{\|{\tilde u}(\mathbf{r},t_{n+1})\|}.
\label{eq-steep-norm}
\end{equation}}
It is interesting to note that \eqref{eq-grad-flow} is commonly referred as the {\em imaginary time evolution} equation, since the right-hand side corresponds to the stationary Gross-Pitaevskii equation. The gradient flow equation (\ref{eq-grad-flow}) (or the related {\em continuous gradient flow} equation, see \cite{bao-norm}) can be viewed as a heat equation in complex variables and, consequently, solved by different classical time integration schemes (Runge-Kutta-Fehlberg \cite{perez-gc,perez-gcM}, backward Euler \cite{bao-norm,ADU,bao2006,bao2008}, second-order Strang time-splitting  \cite{bao-norm,ADU}, etc.). We describe in the following the combined Runge-Kutta-Crank-Nicolson scheme that was successfully used in \cite{dan-2003-aft,dan-2004-aft,dan-2005} to compute stationary three-dimensional BEC configurations for different trapping potentials.

If we write \eqref{eq-grad-flow} under
the general form
\begin{equation}
\frac{\pl u}{\pl t} = {\cal N}(u) + {\cal L}(u),
\label{eq-HL}
\end{equation}
with ${\cal N}(u)$ containing non-linear terms and ${\cal L}(u)$ linear terms, a  combined three-step Runge-Kutta  and  Crank-Nicolson scheme reads: \cite{spalart91,orlandi_book}:
\begin{equation}
    {\frac{u_{k+1}-u_k}{\delta t}}= \underbrace{a_k {\cal N}(u_k)
+b_k {\cal N}(u_{k-1})}_{Runge-Kutta}
+\underbrace{\frac{c_k}{2}{\cal L} \left({u_{k+1}+u_k}\right)}_{Crank-Nicolson},
\end{equation}
where $k=1,2,3$ are the substeps needed to advance the solution from $t_n$ to $t_{n+1}$.
The following values for the coefficients
\begin{eqnarray}
a_1={8\over 15},&\ds a_2={5\over 12},& a_3={3\over 4},\\
   b_1=0,&\ds b_2=-{17\over 60},& b_3=-{5\over 12},\\
      c_1={8\over 15},&\ds c_2={2\over 15},& c_3={1\over 3}.
\end{eqnarray}
ensure the third-order accuracy in time for the Runge-Kutta part and second-order overall accuracy. Note that the intermediate integration time values are $t_k=t_n + c_k \delta t$, with $c_1+c_2+c_3=1$. An important computational advantage is that the scheme is low-storage and self-starting. Indeed, since $b_1=0$ the storage of the solution $u^{n-1}$ at the previous time-step is not necessary.
For numerical purposes, the equation to solve is written as:
\begin{equation}\label{eq-lap-nlint-d2cn}
    \left(\frac{1}{\delta t}-\frac{c_k}{2}{\cal L}\right){q_k}= \left[a_k {\cal N}(u_k)
+b_k {\cal N}(u_{k-1})\right]
+ c_k {\cal L}(u_n), \quad q_k = u_{k+1}-u_k,
\end{equation}
with the variational formulation: find $q_k \in H^1_0({\cal D}, \C)$ such that $\forall v \in H^1_0({\cal D}, \C)$,
\begin{equation}\label{eq-lap-nlint-dvcn}
  \int_{\cal D} \left[\frac{1}{\delta t} q_k v -\frac{c_k}{2}{\cal L}(q_k) v\right] =
  \int_{\cal D}  \left(a_k {\cal N}(u_k)
+ b_k {\cal N}(u_{k-1})\right)v
+ c_k {\cal L}(u_k) v.
\end{equation}
Depending on the choice of the linear operator in (\ref{eq-HL}), we can distinguish between different schemes. In \cite{dan-2003-aft,dan-2004-aft,dan-2005} the linear operator was defined in the classical way: ${\cal L}(u) = \nabla^2(u)$. We use in the following a different choice that resulted in a better stability of the scheme:
\begin{eqnarray} \hspace{0.2cm}
{\cal L}(u) &=& \nabla^2(u) + 2 i \Omega A^t \nabla u,\\
{\cal N}(u) &=& - 2 \left[ g |u|^2 + V_{trap} \right] u.
\label{eq-RK-nonl}
\end{eqnarray}
For this method, the mass conservation constraint \eqref{eq-u-norm} is taken into account by using the discrete normalization \eqref{eq-steep-norm}.

%*******************************************************************************
%%%%%%%%%%%%%%%%%%%%%%%%%%%%%%%%%%%%%%%%%%%%%%%%%%%%%%%%%%%%%%%%%%%%%%%%%%%%%%%
\subsection{Direct minimization: Sobolev gradient descent method}\label{subsec-sobolev}
%%%%%%%%%%%%%%%%%%%%%%%%%%%%%%%%%%%%%%%%%%%%%%%%%%%%%%%%%%%%%%%%%%%%%%%%%%%%%%%

Another method to compute stationary BEC states is to directly minimize the GP energy \eqref{eq-u-energ} using steepest descent methods. It is interesting to note that in the descent method (\ref{eq-grad-flow}), the right-hand side represents the $L^2$-gradient (or {\em ordinary} gradient) of the energy functional.
\Blue{An important improvement of the convergence rate of the descent method is obtained by replacing  the ordinary gradient with the gradient defined on the Sobolev space $H^1({\cal D}, \C)$. The reason is that the use of Sobolev gradients is equivalent to a preconditioning of the ordinary gradient method.
The idea of introducing the Sobolev gradient in a descent method was developed by J. W. Neuberger in the 1970's and is now used in several fields of numerical analysis (see \cite{neuberger-book}). On the related topic of finding minima of the Ginzburg-Landau energy functional for superconductors \cite{du-1992-SIAMREV,du-1992-PRB}, the Sobolev gradient method was first presented in \cite{neuberger-1998-GL}. Recent developments of the method in a finite element setting include the minimization of Schr{\"o}dinger \cite{raza2009-1} or Ginzburg--Landau type functionals \cite{raza2010}.
}

\Blue{
In the framework of computing critical points of the Gross-Pitaevskii energy with rotation, a descent method based on the $H^1$ Sobolev gradient was used in \cite{perez-gc,perez-gcM}, in conjunction with a spectral method for the spatial discretization.
In \cite{dan-2009-par} we have equipped the Sobolev space $H^1$ with a new inner scalar-product and used the associated gradient  to improve the convergence  of the descent method for high rotation frequencies. The new inner product is}
\begin{equation}
\langle u , v \rangle_{H_A} = \int_{\cal D} \langle u , v \rangle  + \langle \nabla_A u , \nabla_A v \rangle,
\label{eq-innp-HA}
\end{equation}
where $\nabla_A = \nabla + i \Omega A^t$, and $\langle \cdot , \cdot \rangle$ denotes the complex inner product. The new Hilbert space is denoted by $H_A({\cal D}, \C)$. Hence, the $H_A$ gradient of the energy functional satisfies the equation:
\begin{equation}
  <\nabla_{H_A} E, v > _{H_A}  = <\nabla_{L^2} E, v > _{L^2}.
  \label{eq-inn-equiv}
\end{equation}

The numerical method introduced in \cite{dan-2009-par} consists in the following steps:
\begin{itemize}
  \item We first compute the gradient $\nabla_{H_A} E$.
Observing  that an equivalent definition of the $H_A$ scalar product is:
\begin{equation}
\label{eq-innp-HA2}
<u, v> _{H_A} = \int_{\cal D} \langle \left[ 1 + \Omega^2 (y^2+x^2) \right]  u , v \rangle + \langle \nabla u , \nabla v \rangle -2i\Omega \langle A^t\nabla u , v \rangle,
\end{equation}
we infer that the gradient ${\cal G} = \nabla_{H_A} E$ could be directly  computed from \eqref{eq-inn-equiv} as the solution of the variational problem:
\begin{equation}
\label{eq-grad-HA}
\int_{\cal D}  \left[ 1 + \Omega^2 (y^2+x^2) \right]  {\cal G}\, v +  \nabla {\cal G}  \nabla v -2i\Omega  (A^t\nabla {\cal G}) v =  \mathrm{RHS}, \quad \forall v \in H^1_0({\cal D}, \C),
\end{equation}
where the right-hand-side term represents the $L^2$ gradient (in the weak form):
\begin{equation}
\mathrm{RHS} = \int_{\cal D}  \nabla u  \nabla v  + 2\left[ V_{trap}\, u + (g |u|^2 )u  - i \Omega  A^t\nabla u \right] v.
\label{eq-grad-L2}
\end{equation}

\item In order to satisfy the mass constraint \eqref{eq-u-norm}, we project the gradient $\nabla_{H_A} E$ onto the {\em tangent space} associated to the constraint. In our case, we project onto the null space of $\beta'(u)$, where $\beta(u)=\int_{\cal D} |u|^2$. The final expression (see \cite{dan-2009-par} for details) of the projection that will be used for numerical implementation is:
\begin{equation}
  P_{u,H_A} {\cal G} = {\cal G} - \frac{\Re \langle u, {\cal G} \rangle_{L^2}}{\Re \langle u, v_{H_A} \rangle_{L^2}}\, v_{H_A},
  \label{eq-projG}
\end{equation}
with $\Re$ denoting the real part, and $v_{H_A}$ computed such as that
\begin{equation}
  \Re \langle v_{H_A} , v \rangle_{{H_A}} = \beta'(u) v = \Re \langle u, v \rangle_{L^2}.
  \label{eq-vX}
\end{equation}

\item The solution is finally advanced following the general descent method:
    \begin{equation}
      u^{n+1} = u^n - \delta t \, P_{u,H_A} {\cal G}(u^n).
      \label{eq-steep-dt}
    \end{equation}
\end{itemize}
It should be noted that the projection step ensures that the norm of the initial condition ($u^0$) is preserved through the iterative process \eqref{eq-steep-dt}.

%%%%%%%%%%%%%%%%%%%%%%%%%%%%%%%%%%%%%%%%%%%%%%%%%%%%%%%%%%%%%%%%%%%%%%%%%%%
\section{Finite element spatial discretization and mesh adaptivity}\label{sec-FE}
%%%%%%%%%%%%%%%%%%%%%%%%%%%%%%%%%%%%%%%%%%%%%%%%%%%%%%%%%%%%%%%%%%%%%%%%%%%

The finite element implementation uses the free software FreeFem++ \cite{freefem}, which proposes a large variety of triangular finite elements (linear and quadratic Lagrangian elements, discontinuous $P^1$, Raviart-Thomas elements, etc.) to solve partial differential equations (PDE) in two dimensions (2D). FreeFem++ is an integrated product with its own high level programming language with a syntax close to mathematical formulations.
FreeFem++ was recently used to test algorithms for the minimization of Schr{\"o}dinger or Ginzburg--Landau functionals \cite{raza2009-1,raza2010}.

%%%%%%%%%%%%%%%%%%%%%%%%%%%%%%%%%%%%%%%%
\subsection{FreeFem++ implementation}
%%%%%%%%%%%%%%%%%%%%%%%%%%%%%%%%%%%%%%%%

It is very easy to implement the variational formulations associated to the above described algorithms using FreeFem++. We outline here  the main features of the finite element implementation that were helpful in writing efficient FreeFem++ scripts. Let ${\cal T}_h$ be a family of triangulations of the domain $\cal D$. We assume that ${\cal T}_h$ is a regular family in the sense of Ciarlet \cite{ciarlet-1978}, with $h>0$ belonging to a generalized sequence converging to zero. We denote by $P^l(T)$ the space of polynomial functions on triangles  $T\in {\cal T}_h$, of degree not exceeding $l\geq 1$ . We also introduce the finite element approximation spaces:
\begin{equation}
  W_h^l = \left\{ w_h \in C^0({\bar {\cal D}_h}); w_h |_T \in P^l(T), \forall T\in {\cal T}_h\right\},
\end{equation}
and
\begin{equation}
  V_h^l = \left\{ w_h \in W_h^l; w_h |_{\Gamma_h} = 0\right\}.
\end{equation}
The finite dimensional space $V_h^l$ is a subspace of $H^1_0({\cal D})$ and therefore will be used to discretize the variational formulations previously written. We use in the following $P^1$ ($l=1$, piecewise linear) finite elements to approximate the solution and a $P^4$ representation of the nonlinear terms. It is interesting to note that  FreeFem++  allows to switch to $P^2$ ($l=2$, piecewise quadratic) finite elements by a simple change of the definition of the generic finite-element space $W_h^l$.

An efficient implementation of the algorithms described in the previous section is obtained using the pre-computation of the complex matrices associated to linear systems. For the imaginary time propagation method, the integral form \eqref{eq-lap-nlint-dvcn} leads to the following linear system:
\begin{equation}
  \left[\frac{1}{\delta t} A_M + \frac{c_k}{2} A_G - \frac{c_k}{2} A_\Omega\right] Q_k = A_M^4. (a_k N_k + b_k N_{k-1}) - {c_k} A_G U_k + {c_k} A_\Omega U_k,
  \label{eq-matr-RK}
\end{equation}
where $U_k$ is the solution vector at substep $k$ of the Runge--Kutta method and $Q_k = U_{k+1}-U_k$. Denoting by  $w_h^l$ the basis functions of the space $V_h^l$,  the matrices in \eqref{eq-matr-RK} are defined in the classical way using $l=1$:
\begin{eqnarray}
(A_M)_{m,p} &=& \int_{{\cal D}_h} (w_h^1)_m (w_h^1)_p, \\
(A_G)_{m,p} &=& \int_{{\cal D}_h} \nabla(w_h^1)_m \nabla(w_h^1)_p, \\
(A_\Omega)_{m,p} &=& (2 i\Omega) \int_{{\cal D}_h} (A^t\nabla (w_h^1)_p) (w_h^1)_m.
\end{eqnarray}
Nonlinear terms $N_k$, corresponding to \eqref{eq-RK-nonl}, are computed with higher accuracy using $P^4$ finite elements. The (non squared) matrix $A_M^4$ is consequently computed as:
\begin{equation}
(A_M^4)_{m,p} = \int_{{\cal D}_h} (w_h^1)_m (w_h^4)_p.
\end{equation}
Previous two-dimensional integrals are computed using a fifth order quadrature formula. If the imaginary time advancement is conducted with fixed size time step, a further optimization comes from the storage of the three matrices of the linear systems corresponding to each substep of the Runge--Kutta integration procedure.

For the Sobolev gradient method, the discrete form of
\eqref{eq-grad-HA} becomes:
\begin{equation}
  A_S G = A_M^4. N_n + A_G U_n - A_\Omega U_n,
  \label{eq-matr-HA}
\end{equation}
with $N_n$ corresponding to a $P^4$ representation of nonlinear terms $2\left( V_{trap} +  g |u_n|^2\right)u_n$. The matrix $A_S$ of the linear system:
\begin{equation}
\label{eq-grad-HA-mat}
(A_S)_{m,p}= \int_{{\cal D}_h}  \left[ 1 + \Omega^2 (y^2+x^2) \right]  (w_h^1)_m\, (w_h^1)_p +  \nabla(w_h^1)_m \nabla(w_h^1)_p -2i\Omega  (A^t\nabla)(w_h^1)_p (w_h^1)_m,
\end{equation}
is computed by a fifth order quadrature formula. An important computational tine saving is obtained if the matrix  $A_S$ is stored and factorized before the time loop \eqref{eq-steep-dt}.

The last point to emphasize concerning the FreeFem++ implementation is that all previous equations are solved in complex variables. As a consequence, the corresponding matrices also have complex elements. The approach used in \cite{raza2009-1}, based on the separation of the real and imaginary part of the unknown variable, results in considerably larger computational times. Besides, this separation is not possible when computing the $H_A$ gradient from \eqref{eq-grad-HA}.

%%%%%%%%%%%%%%%%%%%%%%%%%%%%%%%%%%%%%%%%
\subsection{Adaptive mesh refinement strategy}
%%%%%%%%%%%%%%%%%%%%%%%%%%%%%%%%%%%%%%%%

Mesh adaptivity by metric control is a standard function offered by FreeFem++.
Details on the ingredients used in the metric mesh adaptation
can be found in \cite{hecht-1996-missi,hecht-2000-ijnmf,hecht-1997-aiaa,george-1998,frey-george-1999,moham-piron-2000}.
The key idea is to modify the scalar product used in an automatic mesh generator to evaluate distance and volume, in order to  construct equilateral elements according to a new adequate metric.  The scalar
product is based on the evaluation of the Hessian $\mathcal{H}$ of the variables of the problem. Indeed, for a $P^1$ discretization of a
variable $\chi$, the interpolation error is bounded by:
\begin{equation}
{\cal E } = |\chi - \Pi_h \chi |_0 \leq c \sup_{T\in \mathcal{T}_{h}} \sup_{x,y,z\in T}   |\mathcal{H}(x)|(y-z) . (y-z)
\label{eq1}
\end{equation}
where $\Pi_h \chi $ is the $P^1$ interpolate  of $\chi$, $ |\mathcal{H}(x)|$ is the  Hessian of $\chi$ at point $x$ after being made positive definite, and $.$ denotes the dot product.
We can infer that, if we generate, using  a Delaunay procedure (e.g. \cite{george-1998}), a  mesh with edges close to the unit length  in the metric  $\mathcal{M} ={|\mathcal{H}| \over (c {\cal{E}})}$, the interpolation error ${\cal E}$ will be equally distributed over the edges $a_i$ of the mesh. More precisely, we have
\begin{equation}
{1 \over c {\cal E}} a_i^T {\cal M } a_i \le 1.
\end{equation}
The previous approach could be generalized for a vector variable $\chi=[\chi_1, \chi_2]$. After computing the metrics $\mathcal{M}_{1}$ and $\mathcal{M}_{2}$ for each variable, we define an metric intersection  $\mathcal{M} = \mathcal{M}_{1} \cap \mathcal{M}_{2}$,
such that the unit ball of $\mathcal{M}$ is included in  the intersection of the two  unit balls  of metrics $\mathcal{M}_{2}$ and $\mathcal{M}_{1}$.
For this purpose, we use the  procedure defined in \cite{frey-george-1999}.
Let $\lambda _{i}^{j}$ and $v_{i}^{j}$, ($i,j=1,2$) be the eigenvalues and
eigenvectors of ${\cal M}_{j}$, $j=1,2$. The intersection metric ($\hat{{\cal M}}$) is defined by
 \begin{equation}
 \hat{{\cal M}}={\frac{{\hat{{\cal M}_{1}}+\hat{{\cal M}_{2}}}}{2}}.
 \end{equation}
 where $\hat{{\cal M}_{1}}$ (resp. $\hat{{\cal M}_{2}}$) has the same
 eigenvectors than ${\cal M}_{1}$, (resp. ${\cal M}_{2}$ ) but with
 eigenvalues defined by:
 \begin{equation}
 \tilde{\lambda _{i}^{1}}=\max (\lambda _{i}^{1},{v_{i}^{1}}^{T}{\cal M}_{2}v_{i}^{1}),\quad i=1,2.
 \end{equation}

FreeFem++ uses mesh generation tools developed in \cite{george-1998,frey-george-1999} with the  novelty that
the Delaunay mesh generation procedure introduces  an extra criterion
to keep the new mesh nodes and connectivity maps unchanged as much as possible
when the prescribed mesh by the new metric is similar to the previous mesh. This  reduces the perturbations introduced when the solution is embedded by interpolation from the old mesh to the new one.

The mesh adaptivity strategy used in this work is based on the fact that the energy of the solution decreases during the computation to attain a plateau corresponding to a local minima (see Fig. \ref{fig-time-evol}).
Since we generally use a convergence criterion \cite{dan-2003-aft,dan-2004-aft,dan-2005,dan-2009-par} based on the relative change of the energy of the solution, $\delta E_n = (E_{n+1}-E_n)/E_n  <\varepsilon_c$, we monitor the same quantity to trigger the mesh adaptivity
procedure following the next algorithm:
\begin{enumerate}
  \item choose a sequence of decreasing values $\varepsilon^i \ge \varepsilon_c$, that represent threshold values for the mesh adaptivity;
  \item set $i=1$;
  \item if $\varepsilon^{i+1} < \delta E_n < \varepsilon^{i}$ and $\delta E_n > \varepsilon_c$, call the mesh adaptivity procedure; \Blue{the solution $u$ is interpolated on the new mesh and normalized to satisfy the unitary norm constraint;}
  \item if step 3 was performed $N_{ad} \geq 1$ times,  increase $i$ to $i+1$. \Blue{Limiting the number of mesh refinements for the same threshold, is necessary since, at step 2, the interpolation on the new refined mesh and the normalization of the solution could lead to an increase of the value of $\delta E_n$.}
\end{enumerate}
As an example, for the computation displayed in Figs. \ref{fig-abrikosov} and \ref{fig-time-evol}, we fixed the convergence threshold  to $\varepsilon_c=10^{-8}$ and mesh refinement threshold values to $\varepsilon \in \{10^{-6}, 5\cdot 10^{-7}, 2.5\cdot 10^{-7}, 10^{-8} \}$. The number of calls for the mesh refinement procedure was $N_{ad}=3$ for each fixed threshold. We can notice in Fig. \ref{fig-time-evol} the jump in the energy evolution when the mesh refinement was applied, resulting in a faster convergence to the final value of the energy.

An essential question that remains when defining the mesh refinement procedure is the choice of the mesh adaptivity variable $\chi$. Since vortices are characterized by small cores in which the atomic density rapidly decreases to zero in the vortex center, it may appear obvious to use as mesh refinement variable $\chi=|u|$, the modulus of the wave function. We prove by extensive numerical tests described in the next section  that this approach is not always successful. In exchange, the adaptivity strategy considering simultaneously the real and imaginary part of the solution to compute the metrics, {\em i.e.} $\chi=[u_r, u_i]$, proved effective in capturing the right solution with an important reduction of the computational time compared to fixed mesh calculations. This strategy was applied in computing the complex vortex configuration displayed in Fig. \ref{fig-abrikosov}.

%%%%%%%%%%%%%%%%%%%%%%%%%%%%%%%%%%%%%%%%%%%%%%%%%%%%%%%%%%%
\section{Numerical experiments} \label{sec-numerics}
%%%%%%%%%%%%%%%%%%%%%%%%%%%%%%%%%%%%%%%%%%%%%%%%%%%%%%%%%%

In computing stationary states of rotating Bose-Einstein condensates, the initial condition  $u_0$ plays a crucial role. It was theoretically proved in \cite{jackson-2006-bar} that in a real-time evolution of the rotating condensate, the number of vortices  attained by the condensate depend upon the
rotation history of the trap and on the number of vortices present in the condensate initially. This observation also holds for the imaginary-time evolution: for the same rotation frequency, different stationary states, characterized by closed values of the energy, could be obtained starting from different initial conditions.

Three types of initial conditions are generally used for computing stationary states in a rotating BEC: {\em (i)} condensate without vortices, with a  wave function distribution derived from a physical model, called the Thomas-Fermi approximation; {\em (ii)} condensate described by the Thomas-Fermi model on which vortices could be artificially superimposed using an mathematical ansatz; {\em (iii)} initial state set equal to a converged state for a different rotation frequency $\Omega$ or a different interaction constant $g$. The computation depicted in Figs. \ref{fig-abrikosov} and \ref{fig-time-evol} was performed  for $g=5000$ and started from a converged state obtained for $g=2000$.

The Thomas-Fermi approximation consists in neglecting the contribution of the kinetic energy in the strong interaction regime (large values of $g$). A simplified energy functional is obtained:
\begin{equation}
E_{\TF} (\rho) = \int_{\cal D} V_{trap} |u|^2 + \frac{g}{2} |u|^4,
\end{equation}
with a minimizer corresponding to the so-called Thomas-Fermi atomic density:
\begin{equation}
\rho_{\TF}(r) = |u|^2 =\left(\frac{\mu - V_{trap}}{g}\right)_+,
\label{eq-rhoTF}
\end{equation}
where $\mu$ is the chemical potential. Since $\mu$ is a Lagrange multiplier,
a relation that allows to compute $\mu$ is obtained by imposing
the mass constraint  in \eqref{eq-rhoTF}. The initial condition is finally set to $u_0(x,y)=\sqrt{\rho_{\TF}(x,y)}$.

This model is also useful in estimating the necessary size of the computational domain. When a rotation $\Omega$ is applied, the Thomas-Fermi approximation \eqref{eq-rhoTF} stands with $V_{trap}$ replaced by:
\begin{equation}
V_{eff}(r)= V_{trap}(r) - \frac{ \Omega^2 r^2}{2}.
\label{eq-veff}
\end{equation}
The resulting radius $R_{\TF}^\Omega$, corresponding to the point where $\rho_{\TF}^\Omega=0$, is
used to estimate the size $r_D$ of the domain $\cal D$ in simulations ($r_D > R_{\TF}^\Omega$) .

Initial conditions with vortices are obtained by superimposing to the Thomas-Fermi wave function distribution a simple ansatz for the vortex \cite{dan-2003-aft,dan-2004-aft,dan-2005}.
For example, an initial condition with $N_v$ vortices of radius $\epsilon_v$ and centers $(x_v^i, y_v^i), i=1,\ldots,N_v$ is obtained by imposing
\begin{eqnarray}
 u_0(x,y)=\sqrt{\rho_{\TF}(x,y)}\,  \prod_{i=1}^{N_v} u_{v}^i(x,y), \\
 u_{v}^i(x,y)=\sqrt{0.5
\left\{1+tanh\left[\frac{4}{\epsilon_v}\left(r_l-\epsilon_v\right)\right]\right\}}
\, \exp(i \theta_l),
\end{eqnarray}
where $(r_l,\theta_l)$ are  polar coordinates in the framework centered at $(x_v^i, y_v^i)$. Note that the ansatz is written for singly quantized vortices (winding number equal to 1).

We present in the following different types of numerical experiments. We start with test cases reflecting two different imaginary time evolutions:
{\em (i)}  the number of vortices at convergence remains the same as in the initial condition; {\em (ii)}  new vortices enter the condensate. These experiments will serve to test different strategies for mesh adaptivity and to ascertain the computing time gain offered by the present method. Finally, the method is used to compute complex configurations relevant for physical rotating condensates.

We also mention that the converged final state is characterized by its energy $E(u)$ and angular momentum $L_z(u)$ which gives a measure of the rotation:
\begin{equation}
   L_z(u) = \int_{\cal D} \Re \left(i u^* (A^t \nabla) u \right).
  \label{eq-LZ}
\end{equation}

%%%%%%%%%%%%%%%%%%%%%%%%%%%%%%%%%%%%%%%%%%%%%%%%%%
\subsection{Numerical experiment 1}
%%%%%%%%%%%%%%%%%%%%%%%%%%%%%%%%%%%%%%%%%%%%%%%%%

In laboratory experiments, the condensate is typically confined by a harmonic trapping potential $V_{trap}=r^2/2$. It is easy to see from \eqref{eq-veff} that this potential sets a upper bound for the rotation frequency, since for $\Omega=1$ the centrifugal force balances the trapping force and the confinement of the condensate vanishes. To overcome this limitation, different forms of the trapping potential are currently studied, experimentally and theoretically. We use in this experiment a combined harmonic-plus-quartic potential  \cite{kasamatsu-giant,dan-2004-aft,dan-2005,rot-fetterS} that allows high rotation frequencies.

We set the following parameters of the simulation
\begin{equation}
g = 500, \quad V_{trap}=r^2/2+r^4/4, \quad \Omega = 2.
\label{eq-pot-quartic}
\end{equation}
The computational domain is circular of radius $R_{max}= 1.25 \cdot R_{\TF}^\Omega$, where the Thomas-Fermi radius is for this case $R_{\TF}^\Omega = 3.4$. The initial mesh is generated using $M=200$ equally distributed points on the border of the domain.

\begin{figure}[!h]
\centering
\includegraphics[width=0.75\columnwidth]{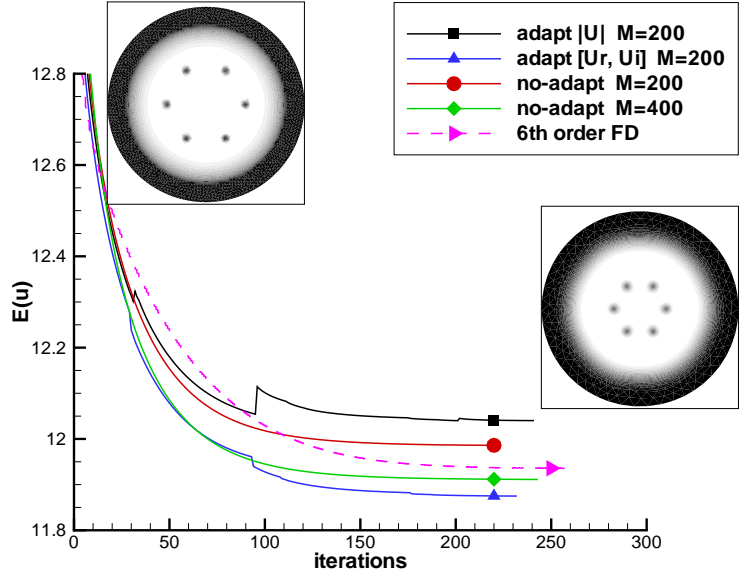}
\caption{\Blue{Computation for  $g=500, \Omega=2$ and combined harmonic-plus-quartic trapping potential. Initial condition with 6 vortices artificially placed at $0.5\cdot R_{\TF}^\Omega$. Energy evolution for constant mesh and different adaptive mesh computations; the result obtained with a 6th order finite difference method is also plotted for comparison.  Density contours ($|u|$) for initial and converged solution (low density in black).}}
\label{fig-exp1-energ}
\end{figure}

The computation is depicted in Fig. \ref{fig-exp1-energ}. The initial condition contains an array of 6 singly quantized vortices equally distributed on the circle of radius $0.5\cdot R_{\TF}^\Omega$. The converged state contains the same number of vortices, but with larger cores than initially set, and placed closer to the center of the condensate, at $0.33\cdot R_{\TF}^\Omega$. Two computations with fixed mesh ($M=200$ and $M=400$) are run and compared to adaptive mesh computations using as adaptivity variable  $\chi=|u|$ and $\chi=[u_r, u_i]$, respectively. Convergence test is set to $\delta E_n \leq 2\cdot 10^{-6}$ for all computations and threshold values for  mesh refinement are chosen as $\varepsilon \in \{ 10^{-2}, 10^{-3},  10^{-4}, 10^{-5}, 10^{-6} \}$. Three mesh refinements are done for each threshold ($N_{ad}=3$).

\Blue{It is interesting to note from Fig. \ref{fig-exp1-energ} the monotone decrease of the energy which is typical for the steepest descent method. This evolution is not affected by the projection method for the unitary norm constraint, as showed for the computations using fixed meshes.
The mesh refinement results in a jump in the energy evolution curve at adaptivity thresholds. As already stated, this is the consequence of the interpolation on the new refined mesh and the normalization of the interpolated solution. Such jumps are naturally less visible close to the convergence, when small variations of the energy are monitored.}

\Blue{In order to assess for the correct behavior of the numerical system, we also compare present finite element results with those obtained using a high order finite difference method. For this purpose, the imaginary time-propagation method presented in section \ref{subsec-rk} was implemented using for the spatial discretization a 6th order compact finite difference scheme that offers spectral-like accuracy \cite{lele-compact}. The method has similarities with that used in \cite{dan-2003-aft,dan-2004-aft,dan-2005} to compute stationary vortex states in a three-dimensional BEC. The finite difference method uses a squared computational domain of size $2 R_{max}$ and a uniform mesh of $105 \times 105$ grid points. The constant mesh size $\delta x = \delta y =0.08$ thus becomes similar to the minimum edge size of the final refined finite element grid ($h_{min} = 0.08$).}

\begin{figure}[!h]
\centering
\includegraphics[width=0.80\columnwidth]{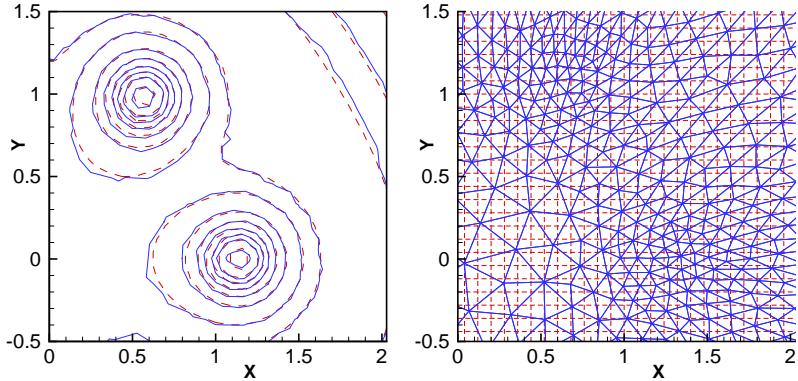}
\caption{\Blue{Computational case depicted in Fig. \ref{fig-exp1-energ}. Comparison between the results obtained with the finite element method for $M=200$ and  mesh adaptivity using $\chi=[u_r, u_i]$ and the 6th order finite difference method with a $105 \times 105$ equally spaced grid. Details of the contours of the atomic density $|u|$  and corresponding grids (dashed lines for the finite difference results).}}
\label{fig-exp1-FEFD}
\end{figure}

\Blue{All computations lead to identical configurations of the final, converged state, as represented in Fig. \ref{fig-exp1-energ}. A detailed comparison between finite element and finite difference results is offered in Fig. \ref{fig-exp1-FEFD}. The finite element grid contains initially 7054 triangles and ends with an adapted mesh with 3722 triangles, while the finite difference mesh has a fixed size of 11025 grid points.
A zoom inside the zone containing two neighboring vortices of the final configuration shows that contours of the atomic density $|u|$ are almost identical. It should be noted that in adapting the finite element mesh, one could impose the values for $h_{max}$ and $h_{min}$, which are the maximum and, respectively, the minimum edge size of the triangular mesh. Reducing the value of $h_{max}$ will result in a finer mesh and smoother contour lines, comparable to those obtained with the high order finite difference discretization.
However, the present comparison is more than satisfactory with a final finite element grid using almost 3 times less grid points than the finite difference setting.}

\begin{table}
\begin{center}
  \begin{tabular}{|l|l|l|l|l|l|c|l|l|l|c|}
  \hline
  &\multicolumn{2}{c}{} &\multicolumn{4}{|c|}{Sobolev gradient method} &\multicolumn{4}{|c|}{Imaginary time method}\\
  \cline{2-11}
Run case   & $M$ & $N_t$ & $E(u)$ & $L_z(u)$ & iter & CPU & $E(u)$ & $L_z(u)$ & iter & CPU \\
    \hline
adapt [$u_r, u_i$] & 200 & 3722  & 11.87 & 5.118 & 232 & 55 & 11.87 & 5.112 & 139 & 54\\
adapt [$|u|$]      & 200 & 2586  & 12.04 & 5.095 & 241 & 44 & 12.02 & 5.088 & 142 & 40\\
no-adapt           & 200 & 7054  & 11.98 & 5.126 & 223 & 72 & 11.91 & 5.085 & 75 & 43\\
no-adapt           & 400 & 27674 & 11.91 & 5.169 & 243 & 315 & 11.83 & 5.125 & 92 & 211\\ \hline
  \end{tabular}

\end{center}
  \caption{$\Omega=2$: run cases corresponding to the numerical experiment depicted in Fig. \ref{fig-exp1-energ}. Parameters of the initial mesh (number of points $M$ placed on the border of the circular domain to generate the mesh and number of triangles $N_t$), energy $E(u)$ and angular momentum $L_z(u)$ of the final state, and computational efficiency (number of iterations and computational $CPU$ time). }
  \label{tab-exp1-HA}
\end{table}

The exact values of the energy $E(u)$ and angular momentum $L_z(u)$ characterizing the final state are shown in Tab. \ref{tab-exp1-HA}. Compared to the fixed mesh computation using a refined mesh ($M=400$), the adaptivity strategy using two variables ($\chi=[u_r, u_i]$)  gives the closest energy value. \Blue{We can also see from Fig. \ref{fig-exp1-energ} that this is also the case when comparing with the 6th order finite difference result.} Meanwhile, this adaptive mesh strategy results in a computational time reduction by a factor of 6 for the Sobolev gradient method and by a factor of 4 for the imaginary time propagation method. Table \ref{tab-exp1-HA} also shows that the two numerical methods used to compute stationary states behave similarly. Since this is also the case for all subsequent numerical experiments,  we discuss in the following, for the sake of simplicity,  only the results obtained with the Sobolev gradient method. This method has also the advantage to allow a constant time step for different mesh densities (see also \cite{dan-2009-par}).

\begin{figure}[!h]
\centering
\includegraphics[width=0.70\columnwidth]{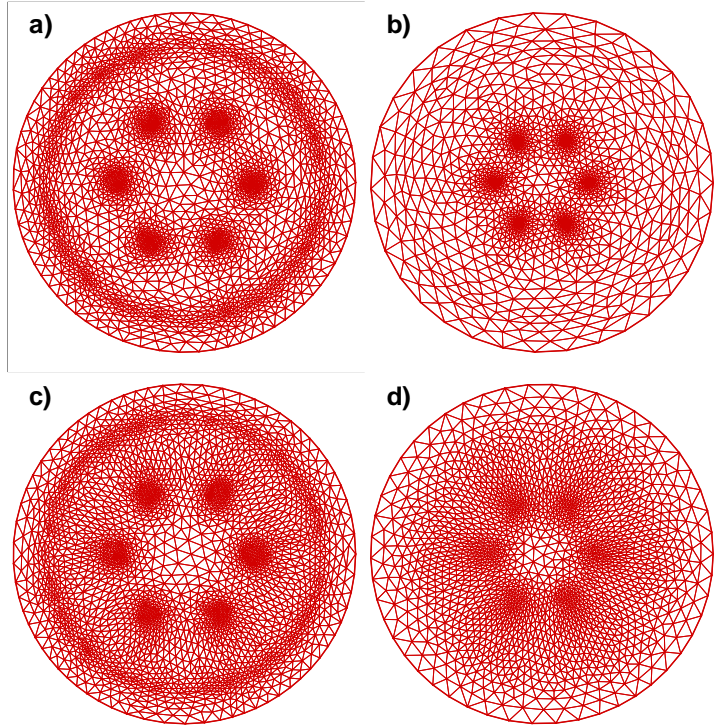}
\caption{Mesh evolution during the computation for experiment 1 (see Fig. \ref{fig-exp1-energ}). First ($\varepsilon=10^{-2}$) and final ($\varepsilon=10^{-5}$) refined meshes are represented for the adaptive mesh strategy using $\chi=|u|$ (a and b) and $\chi=[u_r, u_i]$ (c and d).}
\label{fig-exp1-mesh}
\end{figure}

The mesh evolution for the two adaptivity strategies can be followed  in Fig. \ref{fig-exp1-mesh}. Only meshes for the  first ($\varepsilon=10^{-2}$) and final ($\varepsilon=10^{-5}$) thresholds are represented. It can be easily seen that adaptivity taking into account only the modulus of the solution results in a very dense mesh in the center of vortices. Adaptivity following the real and imaginary part of the solution also generates a refined mesh in the core of vortices, but also a dense mesh from vortices towards the border of the condensate. This allows to have a better representation of the phase of the solution (as previously pointed out when discussing Fig. \ref{fig-abrikosov}). We shall see in the following that this feature is crucial for the success of the adaptivity strategy when more complicated cases are computed.

\pagebreak

%%%%%%%%%%%%%%%%%%%%%%%%%%%%%%%%%%%%%%%%%%%%%%%%%%
\subsection{Numerical experiment 2}
%%%%%%%%%%%%%%%%%%%%%%%%%%%%%%%%%%%%%%%%%%%%%%%%%

\begin{figure}[!h]
\centering
\includegraphics[width=0.80\columnwidth]{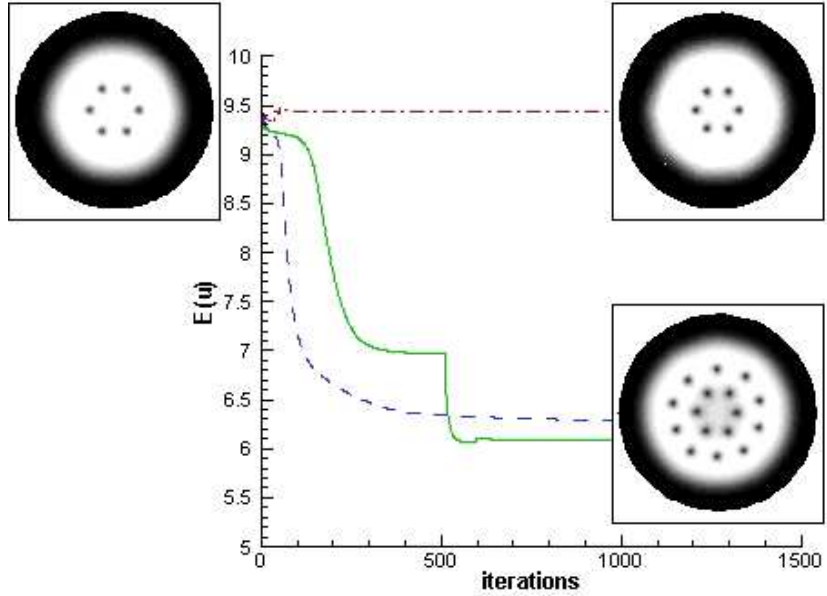}
\caption{Computation for  $g=500, \Omega=2.5$ and combined harmonic-plus-quartic trapping potential. Computations start from the converged state obtained for $\Omega=2$. Energy evolution for constant mesh ($M=400$, dashed line) and different adaptive mesh computations: $\chi=|u|$ (dash-dot line) and $\chi=[u_r, u_i]$ (solid line). Density contours ($|u|$) for initial and converged solution (low density in black).}
\label{fig-exp2-energ}
\end{figure}

In this experiment, we consider the same parameters as for experiment 1 and increase the rotation frequency to $\Omega=2.5$. The initial condition is the converged state previously computed for $\Omega=2$. For this case, new vortices are nucleated inside the condensate and the final state contains a second circle of 10 vortices. Figure \ref{fig-exp2-energ} shows that only the adaptive mesh strategy based on $\chi=[u_r, u_i]$ converges to a similar stationary state as the computation using the fixed refined mesh ($M=400$). This is also
visible from Tab. \ref{tab-exp2-HA}, when comparing the values of the energy and angular momentum of the final state. In exchange,
mesh refinement using $\chi=|u|$ do not allow the nucleation of new vortices; as a consequence, the energy of the system is not decreasing and the final state has the same configuration as the initial condition. It is important to note from Tab. \ref{tab-exp2-HA} that the successful adaptive mesh strategy allows a tremendous (factor of 10) gain of computational time.

\begin{table}
\begin{center}
  \begin{tabular}{|l|l|l|l|l|l|c|l|l|l|c|}
  \hline
  &\multicolumn{2}{c}{} &\multicolumn{4}{|c|}{Initial condition 1} &\multicolumn{4}{|c|}{Initial condition 2}\\
  \cline{2-11}
Run case   & $M$ & $N_t$ & $E(u)$ & $L_z(u)$ & iter & CPU & $E(u)$ & $L_z(u)$ & iter & CPU \\
    \hline
adapt [$u_r, u_i$] & 200 & 8968  & 6.08 & 11.95 & 1266 & 321
                                 & 5.94 & 12.87 & 222  & 195 \\
adapt [$|u|$]      & 200 & 2540  & 9.43 & 5.41  & 456  & 33
                                 & 7.14 & 11.30 & 3280 & 947 \\
no-adapt           & 400 & 27654 & 6.23 & 12.81 & 3041 & 3368
                                 & 6.31 & 13.01 & 249 & 327 \\ \hline
  \end{tabular}
\end{center}
  \caption{$\Omega=2.5$. Same legend as for table \ref{tab-exp1-HA}. Initial condition 1 is the converged state obtained for $\Omega=2$ (Fig. \ref{fig-exp2-energ}) and initial condition 2 contains an artificially generated state with 3 arrays of vortices (Fig. \ref{fig-exp2-bis-energ}).}
  \label{tab-exp2-HA}
\end{table}

\begin{figure}[!h]
\centering
\includegraphics[width=0.65\columnwidth]{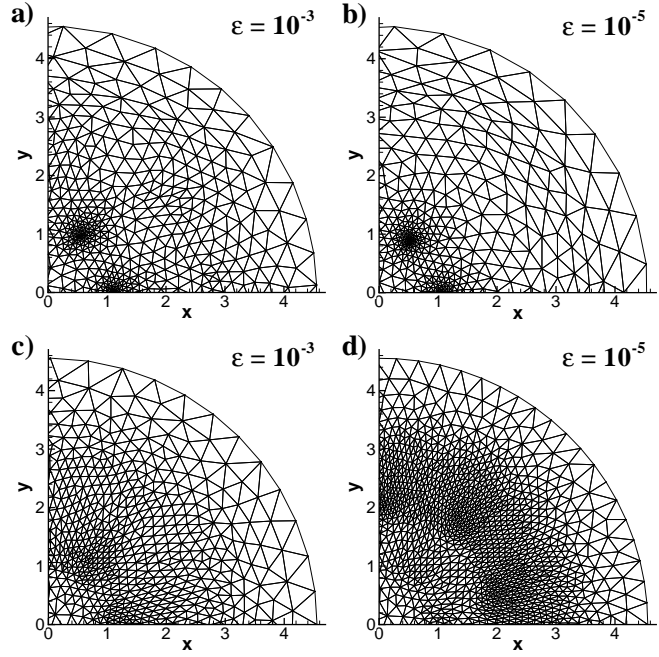}
\caption{$\Omega=2.5$ Mesh evolution during the computation for experiment 2 (see Fig. \ref{fig-exp2-energ}). Refined meshes corresponding to thresholds $\varepsilon=10^{-3}$ and $\varepsilon=10^{-5}$. Adaptive mesh strategy using $\chi=|u|$ (a and b) and $\chi=[u_r, u_i]$ (c and d).}
\label{fig-exp2-mesh}
\end{figure}

The explanation for the failure of the adaptive method based solely on the modulus of the solution is offered in Fig. \ref{fig-exp2-mesh}. A computation is subject to inherent numerical perturbations  that will trigger the nucleation of new vortices. Such perturbations  usually have small amplitudes, and the refinement based on the modulus of the solution will damp them since the mesh size in these regions is not small enough to capture them. The adaptive mesh strategy using $\chi=[u_r, u_i]$ generates refined meshes over larger regions than the core of vortices (see Figs. \ref{fig-exp2-mesh}c and \ref{fig-exp2-mesh}d) and consequently allow the nucleation of new vortices.

\begin{figure}[!h]
\centering
\includegraphics[width=0.80\columnwidth]{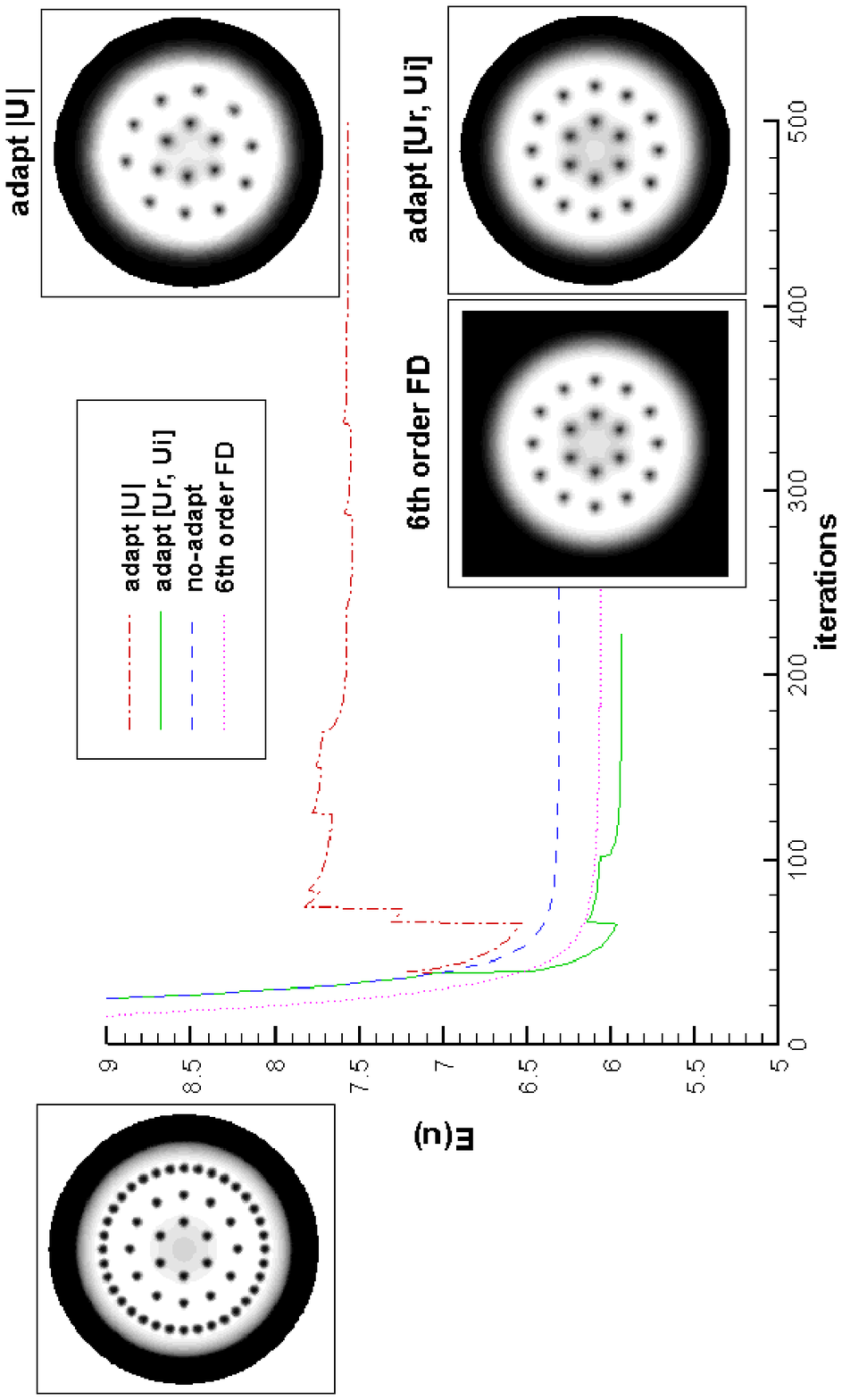}
\caption{\Blue{Computations with the same parameters as in Fig. \ref{fig-exp2-energ}, but starting now from an artificial initial condition containing three arrays of vortices. Only the adaptivity strategy with $\chi=[u_r, u_i]$ allows to obtain a correct final state, which is almost identical to the configuration obtained with a refined fixed mesh ($M=400$) or a 6th order finite difference (FD) method with a $125 \times 125$ uniform grid.} }
\label{fig-exp2-bis-energ}
\end{figure}

An intriguing question that one could raise after analyzing numerical experiments 1 and 2 is whether the adaptive mesh strategy based on the modulus is successful if the perturbation necessary to nucleate vortices are present in the initial condition. This question is addressed by performing computations starting from an initial condition  with three arrays containing 6, 12, and 36 vortices, respectively. The external circle of vortices plays the role of a dense perturbation field that could trigger vortices for this high rotation frequency. Figure \ref{fig-exp2-bis-energ} shows that, once again, only the adaptive mesh strategy considering simultaneously the real and imaginary pert of the solution is successful. \Blue{The converged configuration for this computation is very similar to that obtained when using a refined ($M=400$, $h_{min}=0.0506$) fixed mesh or a 6th order finite difference method using a $125 \times 125$ uniformly spaced grid ($\delta x =\delta y = 0.053$).}


%%%%%%%%%%%%%%%%%%%%%%%%%%%%%%%%%%%%
\subsection{Condensates with giant vortex or dense vortex lattice}
%%%%%%%%%%%%%%%%%%%%%%%%%%%%%%%%%%%%%

In order to assess for the efficiency of our numerical system, we consider in this section two cases closer to experimental configurations. Such cases are difficult to compute since they involve high values for the atomic interaction constant $g$ and/or rotation frequency $\Omega$.

\begin{figure}[!h]
\centering
\includegraphics[width=0.70\columnwidth]{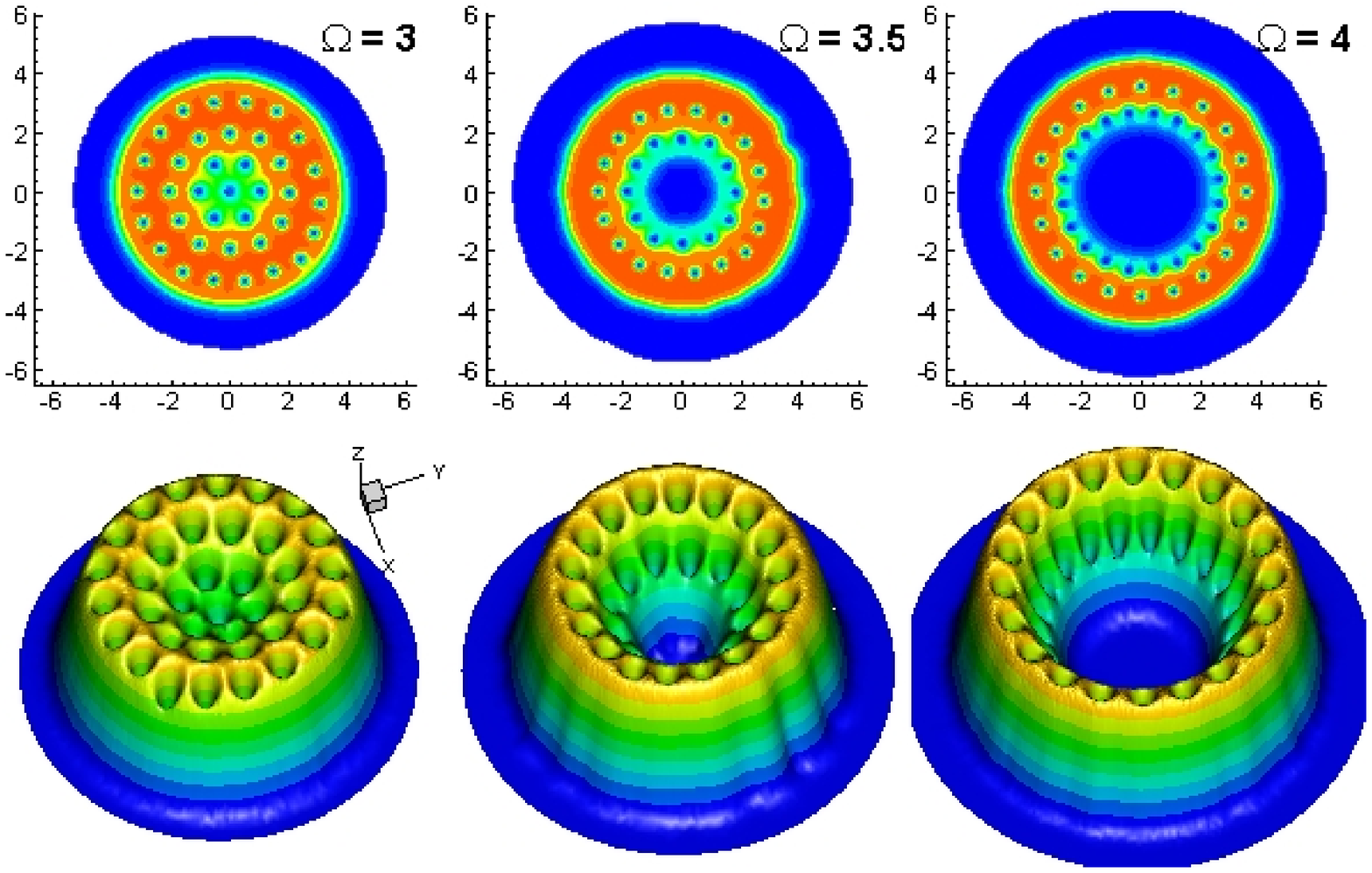}
\caption{Condensate trapped in a harmonic-plus-quartic potential ($g=1000$). Two- and three-dimensional representation of the atomic density contours (low density in black) for increasing values of the rotation frequency $\Omega$. Note the formation of a giant vortex (hole) in the center of the condensate.}
\label{fig-quartic-all}
\end{figure}

The first case considers the condensate trapped in the harmonic-plus-quartic potential \eqref{eq-pot-quartic}, but with higher atomic interaction constant, $g=1000$. Figure \ref{fig-quartic-all} shows the evolution of the stationary state of the condensate when the rotation frequency is increased. Vortices in the center of the condensate progressively merge to form a giant hole, also called giant vortex. This intriguing configuration has been intensively studied in the physical  literature  \cite{kasamatsu-giant,dan-2004-aft,dan-2005,rot-fetterS}. The adaptive mesh refinement is very useful in computing such cases since the atomic density in a large zone in the center of the condensate is close to zero. As a consequence, large triangles are generated in the center of the condensate, while the mesh is highly refined in the annulus zone, where vortices nucleate. For instance, the simulation  for $\Omega=4$ started with an initial mesh with $N_t=18\,670$ triangles and ended with a fine mesh with $N_t=69\,859$ triangles. For reference, a constant mesh that offers a similar mesh density  in the annular zone is obtained for $M=600$ and contains $N_t=108\,212$ triangles, since all the computational domain is finely meshed.

\begin{figure}[!h]
\centering
\includegraphics[width=0.52\columnwidth,angle=-90]{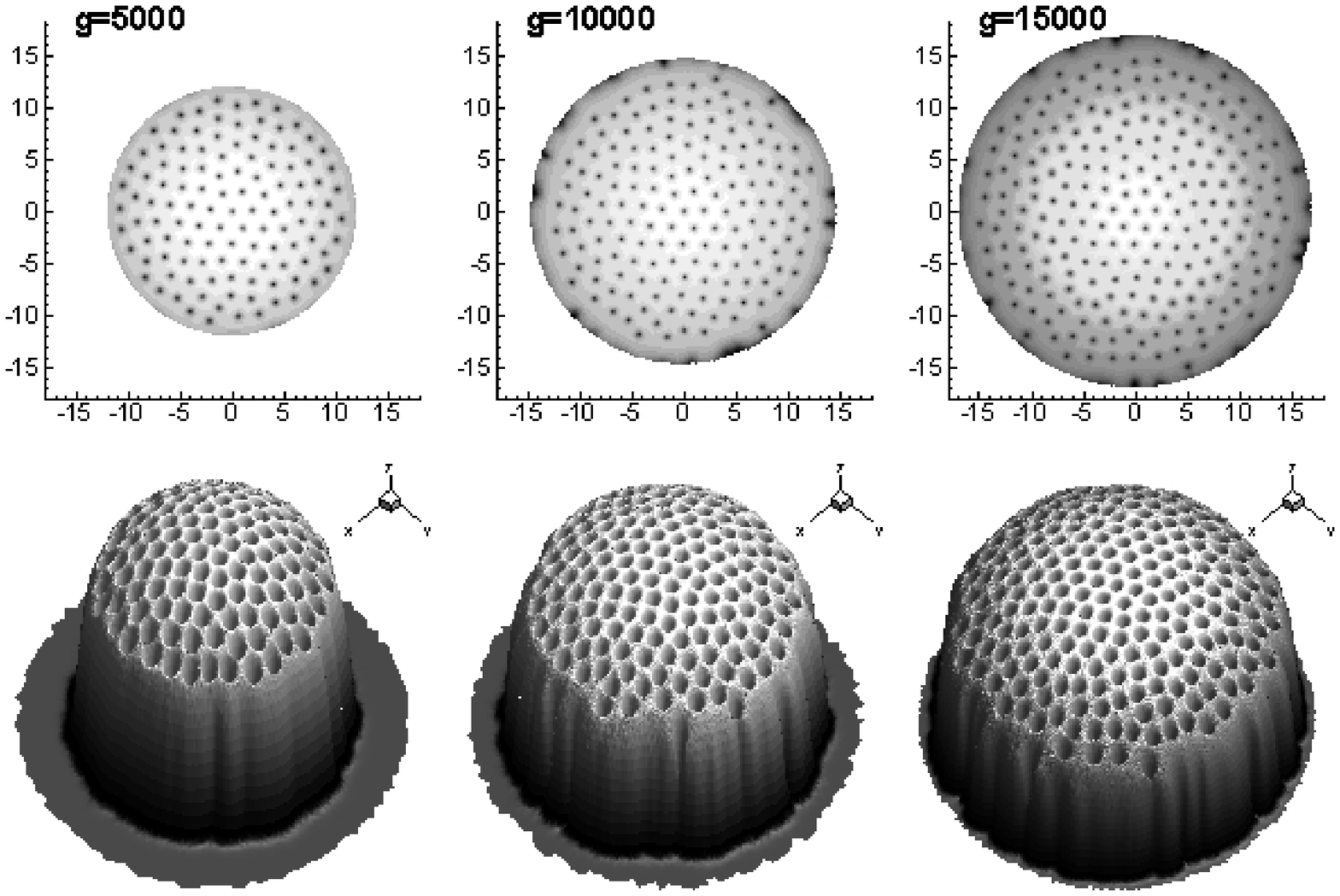}
\caption{Condensate trapped in a harmonic potential ($\Omega=0.95$). Two- and three-dimensional representation of the atomic density contours (low density in black) for increasing values of the atomic interaction constant  $g$. Note the increase of the number of vortices with increasing values of  $g$.}
\label{fig-harmonic-all}
\end{figure}

The second configuration considers the case, displayed  in Figs. \ref{fig-abrikosov} and \ref{fig-time-evol}, of the condensate trapped in harmonic potential and rotating at $\Omega=0.95$. We recall that for this case the rotation frequency cannot exceed $\Omega=1$. The difficulty for this case is to increase the atomic interaction constant $g$ that sets the amplitude of the nonlinear term. Figure \ref{fig-harmonic-all} displays the converged configurations for increasing $g=5000, 10 000$ and $15 000$. The condensate becomes larger with increasing $g$, and, consequently, contains more and more vortices that arrange into a regular triangular lattice (or Abrikosov lattice). The large number of vortices present in the condensate requires refined meshes
making the computations very costly. For reference, the final refined meshes contain, for the three cases, $N_t= 238\,262$, $405\,405$, and, $620\,706$ triangles, respectively. Nevertheless, such computations performed with FreeFem++ remain affordable on a single processor computer.

%%%%%%%%%%%%%%%%%%%%%%%%%%%%%%%%%%%%%%%%%%%%%%%%%%%%%%%%%%%%%%%%%%%%%%%
\section{Summary}
%%%%%%%%%%%%%%%%%%%%%%%%%%%%%%%%%%%%%%%%%%%%%%%%%%%%%%%%%%%%%%%%%%%%%%%

We have shown in this work that  low-order finite element methods with mesh adaptivity are a valid alternative of commonly used high-order methods in computing stationary vortex states of a fast-rotating Bose-Einstein condensate. The mesh refinement using metric control proved effective in computing difficult cases with a large number of vortices or with giant vortex. We showed by extensive numerical tests that adaptive mesh strategy using simultaneously the real and imaginary part of the solution to compute metrics is the successful approach. The strategy based only on the modulus of the solution failed for complicated test cases. An effective algorithm for mesh adaptivity was proposed, with an important computational time reduction over computations using refined fixed meshes.

The present finite element discretization with mesh adaptivity was tested with two numerical methods for computing stationary states: a Sobolev gradient descent method for direct minimization of the energy functional and a method based on the imaginary time propagation of the wave function describing the condensate. The method is, however, of more general interest, and could be used in conjunction with different numerical methods for computing imaginary or real time evolution of superfluid systems with vortices, such as rotating Bose-Einstein condensates or type II superconductors. \Blue{In this context, it is interesting to mention that, after the present manuscript had been completed, the recent review paper \cite{du-2005-GL} was brought to our attention. Among the remaining issues in developing numerical methods for computing vortex states in superconductors, adaptive methods were considered in \cite{du-2005-GL} as challenging because of the complicated patterns of the solution with vortices. The necessity to refine the mesh not only around vortex cores was intuitively recalled when discussing the different patterns displayed by the the real and imaginary parts of the solution. The present study confirms in some sense this intuition and offers an effective method to answer the challenging question of computing solutions with quantized vortices.}

%%%%%%%%%%%%%%%%%%%%%%%%%%%%%%%%%%%%%%%%%%%%%%%%%%%%%%%%%%%%%%%%
% BibTeX users please use
%\bibliographystyle{elsarticle-num}
%\bibliography{../../../ALL_BIBTEX/bib-bec,../../../ALL_BIBTEX/danaila_publis,../../../ALL_BIBTEX/bib-fem,../../../ALL_BIBTEX/bib-vring}

\end{document}